\documentclass[12pt,english,dvips]{scrartcl}
\usepackage{babel}
\usepackage[T1]{fontenc}
\usepackage[ansinew]{inputenc}

\def\beq{\begin{equation}}
\def\eeq{\end{equation}}
\def\munu{{\mu\nu}}
\def\vett{\overrightarrow}
\def\n{\vett{n}}
\def\f{\vett{f}}
\def\e{\vett{e}}
\def\ev{\vett{\eta}}
\def\bitem{\begin{itemize}}
\def\eitem{\end{itemize}}
\def\d{\partial}

\def\di{\d_i}
\def\dj{\d_j}
\def\dm{\d_{\mu}}
\def\dn{\d_{\nu}}

\def\alphabeta{\alpha\beta}
\def\M{\mathcal{M}^4}
\def\5M{\mathcal{M}^5}
\def\S{{\Sigma}^3}
\def\2f{{\phi}^2}
\def\teta{\vartheta}
\def\i{\hat{I}}
\def\j{\hat{J}}
\def\K{{}^4\!K}

\def\ds5{d s_{(5)}}
\def\bear{\begin{array}}
\def\ear{\end{array}}

\def\L{\mathcal{L}}

\begin{document}

\begin{titlepage} \vspace{0.2in} 

\begin{center} {\LARGE \bf 

On the ADM decomposition of the 5-D Kaluza-Klein model}\\ 
\vspace*{1cm}
{\bf Valentino Lacquaniti $\star$ and Giovanni Montani $\diamond$}\\
\vspace*{1cm}
$\star$ Department of Physics "E.Amaldi ", \\ University of Rome "Roma Tre", \\
Via della Vasca Navale 84, I-00146, Rome , Italy \\ e-mail: lacquaniti@fis.uniroma3.it  

 $\diamond$ $\star$ ICRA---International Center for Relativistic Astrophysics\\ 
Dipartimento di Fisica (G9),\\ 
University  of Rome, ``La Sapienza",\\ 
Piazzale Aldo Moro 5, 00185 Rome, Italy.\\ 
e-mail: montani@icra.it\\ 
\vspace*{1.8cm}

%PACS: 11.15.-q, 04.50.+h \vspace*{1cm} \\

{\bf   Abstract  \\ } \end{center} \indent
Our purpose is to recast KK model in terms of ADM variables. We examine and solve the problem of the consistency of this approach, with particular care about the role of the cylindricity hypothesis. We show in details how the KK reduction commutes with the ADM slicing procedure and how this leads to a well defined and unique ADM reformulation. This allows us to consider the hamiltonian formulation of the model and can be the first step for the Ashtekar reformulation of the KK scheme. Moreover we show how the time component of the gauge vector arises naturally from the geometrical constraints of the dynamics; this is a positive check for the autoconsistency of the KK theory and for an hamiltonian description of the dynamics which wants to take into account the compactification scenario: this result enforces the physical meaning of KK model. 

\end{titlepage}

\section{Introduction}
%AAAAAAAAAAAAAAAAAAAAAAAAAAAAAAAAAAAAAAAAAAAAAAAAAAAAAAAAAAAAAAAAAAAAAAAAAAAAAAAAAAAAAAAAAAAAAAAAAAAAAAAAAAAAAAAAAAAAAAAAAAAAAAAAAAAAAAAAAAAAAAAAAAAAAAAAAAAAAAAAAAAAAAAAAAAAAAAAAAAAAAAA
Apart from mathematical difficulties, the main task in the research of a unification theory is to go beyond the conceptual difference we have between gravitation and other fundamental interactions. In the first case, General Relativity provides a space-time geometrical picture of the interaction, while in the other cases unification is achieved within the Yang-Mills scheme by the gauge picture.
Several paths of research attempt to go over this "gauge-geometry" dualism. A first way is to search a suitable reformulation of the General Relativity's dynamics in order to recast gravitational field like a gauge one; a valuable example of this path is the reformulation in terms of "Ashtekar variables" (which leads to loop quantum gravity), that allow us to have a model of gravitation with the symmetry of the SU(2) group, via the use of  the 4-bein formalism and the space-time slicing of the metrics.
A second way is to consider extra-dimensional theories; by the presence of extra degrees of freedom it's possible to have a description of gravitation and others fields in a geometrical unified picture. The idea of multi-dimensions is the basic statement of Kaluza-Klein schemes and it's also common in String and Supergravity theories.
In this work we want to pick up two fundamentals statements of these approaches and to take them into account in the same scheme: the need to make a space-time splitting, in order to have a well-defined time variables, and the idea of the dimensional extension of the general relativity.
Our purpose is to consider the 5-D Kaluza-Klein model and its reformulation in terms of ADM variables.

The original KK model was presented in 1921 (\cite{kaluza}) and 1926 (\cite{klein1},\cite{klein2}) and succesively developed by Jordan\cite{jordan} and Thirry\cite{thirry}; it unified gravitational and electromagnetic field in a five-dimensional model. Actually, the great success of these theories (and the reason of their full development) relies on the formulation of non-Abelian gauge theories, via the geometrization of the Yang-Mills fields (\cite{cianfrani},\cite{montanikk},\cite{modernkk},\cite{cremmer}).
The ADM splitting (\cite{adm}) consists in the reformulation of the dynamics in terms of variables with well-defined properties of transformations under pure spatial diffeomorphisms
and it's  required when we want to consider the hamiltonian formulation of the dynamics (\cite{gravitation},\cite{kuchar}).

As a first objective, the ADM recasting of KK model allows us to consider the hamiltonian formulation of the KK dynamics and its canonical quantization in a  scheme similar to the Wheeler-DeWitt one. As a second objective this work could be viewed as a first step in the Ashtekar reformulation of the KK model.
Third, as we'll see in more details, in our task we have to check the consistency of ADM splitting as far as the KK model is concerned. Indeed, the prize we have to pay in order to achieve the physical meaning of the KK scheme is the breaking of  the 5-D invariance of the theory and this could be not consistent with the hypotheses on which ADM splitting is constructed. Furthermore, in our ADM decomposition we have to consider the gauge (electromagnetic)-field that appears in our dynamics; it appears proper to request that the time component of this field come from the Lagrange multipliers of the model (that are usually revealed by the ADM reformulation), as it happens, for instance, in Supergravity, and as it must be, indeed, if we want to reproduce correctly  ordinary electrodynamics.
Actually, this kind of check represents a striking analysis of the consistency of KK model itself, and this is, indeed, the main part of this work.
We'll prove in details that ADM decomposition is consistent with the KK reduction
 and we'll see that the time component of the gauge field is correctly related to the hamiltonian constraints due to diffeomorphism invariance.

In the next section we'll present a short review of the KK 5-D model and the ADM splitting  procedure; results we'll show are  known in literature, so we'll only want to stress some features of interest for the development of our work. In more detail, about KK model we'll stress how the gauge invariance for electromagnetic field arises as a particular case of diffeomorphism invariance and we'll emphasize how the restrictive hypotheses of KK model  provide a break of the 5-D Poincar\'e group; as far as the ADM splitting is concerned we'll be interested in to see why the hamiltonian formulation require a space-time slicing and how this formulation reveals the existence of a hamiltonian constrained dynamics.
In section (\ref{tre})  the real work starts: we'll face the problem to recast KK model in ADM variables; as told before we'll examine the mathematical and physical consistency of this approach by studying in details the resulting metrics and dynamics.
Once proved the consistency of our procedure, in  section (\ref{quattro}), we'll investigate the role of the  cylindricity condition.  
Finally, in the last section we summarize our results and take a look of the perspectives of this work.   %AAAAAAAAAAAAAAAAAAAAAAAAAAAAAAAAAAAAAAAAAAAAAAAAAAAAAAAAAAAAAAAAAAAAAAAAAAAAAAAAAAAAAAAAAAAAAAAAAAAAAAAAAAAAAAAAAAAAAAAAAAAAAAAAAAAAAAAAAAAAAAAAAAAAAAAAAAAAAAAAAAAAAAAAAAAAAAAAAAAAAAA%
\section{Basic Statement}
\label{due}
\subsection{Kaluza-Klein splitting}
%AAAAAAAAAAAAAAAAAAAAAAAAAAAAAAAAAAAAAAAAAAAAAAAAAAAAAAAAAAAAAAAAAAAAAAAAAAAAAAAAAAAAAAAAAAA%

In the KK 5-D model (also called Abelian model), we consider a curved manifold that is a direct product $\M\otimes S^1$, where $\M$ is the ordinary space-time and $S^1$ is a space-like loop.
The dynamics is linked to the $15$ degrees of freedom of the metric tensor and to the 5-D action:
$$
ds_{(5)}=J_{AB}dx^A dx^B \quad\quad\quad A,B=0,1,2,3,5 
%{}\footnote{This is the historical notation, which emphasizes the role of the fifth dimension}
$$
$$
S=-\frac{1}{16\pi G_{(5)}}\int\!\!d^5\!x \sqrt{J}\,\, {}^5\!R
$$
We notice that we employ the historical notation, which emphasizes the role of the fifth dimension.
We impose two conditions:
\bitem
\item All components of the metrics are independent on the extra coordinate $x^5$ (cylindricity condition)
\item The $J_{55}$ component of the metrics is a scalar.
\eitem
Clearly these conditions break, in the 5-D ambient, the invariance of the dynamics under generic diffeomorphism and at the same time we lose the holding of the equivalence principle.
This is known in literature as a "spontaneous compactification scenario" (\cite{salam}). Indeed we have to consider a restriction of allowable diffeomorphisms in our theory: the proper class under which the dynamics  defined by the above conditions is unchanged is set as follows (we'll denote these as KK diffeomorphism):
\beq
\left\{
\begin{array}{l}
x^5=x^{5'}+ek\Psi(x^{\mu'}) \\
\\
x^{\mu}=x^{\mu}(x^{\nu'})
\end{array}
\right.
\label{KKtrasf}
\eeq 
where $\Psi$ is a scalar functions and $ek$ an appropriate dimensional constant ($ek=\frac{\sqrt{4G}}{c^2}$). As we can see
 the transformations for the ordinary space-time are free, and this allow us to restore general relativity, while for the extra dimension we only admit  translations. This symmetry breaking
is principally due to the cylindricity conditions, which sets the existence of a particular killing vector, i.e. $(0,0,0,0,1) $, or, in other words, sets the existence of a preferred direction in the 5-D space.
However,it is only by this breaking that we can take into account at the same time the gravitational field and a gauge field like the electromagnetic one.
In fact, the reduction of the allowable transformation allows us to consider the following reformulation of the 5-D metric tensor:
\beq
\left\{
\begin{array}{l}
J_{55}=-\phi^2 \\
J_{5\mu}=-\phi^2(ek)A_{\mu}  \quad\quad\quad \mu,\nu=0,1,2,3 \\
J_{\munu}=g_{\munu}-\phi^2 A_{\mu}A_{\nu}
\end{array}
\right.
\label{splittingKK}
\eeq
In these formulas , respect to KK diffeomorphisms, $\phi$ is a scalar field, $g_{\munu}$ is a 4-D tensor and represents the ordinary 4-D metric tensor, and $A_{\mu}$ transforms in the following way:
$$
A_{\mu}'=A_{\nu}\frac{\d x^{\nu}}{\d x^{{\mu}'}}+\frac{\d\Psi}{\d x^{{\mu}'}}
$$
So $A_{\mu}$ is a 4-d vector containing an additional gauge-part. For translations along the fifth dimension, $A_{\mu}$ only transforms  like a gauge vector in a flat space, while for ordinary space-time diffeomorphism transforms like a 4-D vector in a curved manifold. This shows how the gauge invariance is a particular case of the more general invariance for diffeomorphism.
Finally, we have to examine the action that is naturally split in consequence of the metrics reduction.
According to the above reformulation (\ref{splittingKK}) we have the following splitting of the 5-D curvature scalar:
\beq
{}^5\!R=R+\frac{2}{\phi}g^{\munu}\nabla_{\mu}\d_{\nu}\phi+\frac{1}{4}(ek)^2\phi^2F_{\munu}F^{\munu}
\label{rsplittingkk}
\eeq
where $R$ is the usual curvature scalar and $F_{\munu}=\dm A_{\nu}-\dn A_{\mu}$.
Now we can put this expression in the 5-D action and use the hypothesis of closed extra-dimension and cylindricity condition to make a dimensional reduction. We can integrate over $d x^5$ and define the usual constant $G$ as follows:
$$
\frac{1}{16\pi G}=\frac{1}{16\pi G_{(5)}} \int\!d x^5
$$
Finally, by observing that we have $\sqrt{J}=\phi\sqrt{-g}$, after KK splitting the action reads:
\beq
S=-\frac{1}{16\pi G} \int\!d^4\!x \sqrt{-g}\, (\phi R+2\nabla_{\mu}\d_{\nu}\phi+\frac{1}{4}(ek)^2\phi^3F_{\munu}F^{\munu} )
\label{kkaction}
\eeq
Hence our model describes the coupling of the  gravitation field with a gauge field that we interpret as the electromagnetic one plus a scalar field $\phi$. The whole dynamics is invariant under KK diffeomorphisms or, it's the same, under space-time diffeomorphisms plus a gauge field transformation.
If the scalar field were set as a constant (as an additional hypothesis of the theory), the above action would reproduce exactly the Einstein-Maxwell theory. However we can admit this field in the model by simply requiring that, at least at the present cosmological era, $\phi$ be quasi-static, so that its effect would be unobservable. In the original theory $\phi$ was taken as constant, but in modern researches its presence is allowed in the model and it can be viewed as a massless boson of zero spin field (\cite{cianfrani},\cite{modernkk},\cite{witten}). In any case, regardless of $\phi$, holds the most valuable feature that, we repeat , gauge-invariance is ruled out as a particular case of the most general diffeomorphism invariance, as a consequence of the symmetry breaking in the 5-D ambient.
\subsection{ Rules of ADM splitting}
%AAAAAAAAAAAAAAAAAAAAAAAAAAAAAAAAAAAAAAAAAAAAAAAAAAAAAAAAAAAAAAAAAAAAAAAAAAAAAAAAAAAAAAAAAAAAAAAA
Now, for sake of simplicity, we'll examine the rules of ADM splitting for a 4-D manifold (i.e. the usual space-time) but they can be easily extended for every number of dimensions. So let's consider a $\M$ manifold, with tensor metric $g_{\munu}$ and internal coordinates $u^{\mu}$ ($\mu=0,1,2,3$). At a fixed time we can  identify a spatial 3-D hypersurface $\S$ by a parameterization like $u^{\mu}=u^{\mu}(x^i)$ where $x^i$ ($i=1,2,3$) define local coordinates on $\S$ (\cite{kuchar},\cite{kolb},\cite{montaniadmtempo}). The possibility to identify always such a hypersurface is due to a theorem of Geroch\cite{geroch} that holds for homogeneous and isotropic manifold of hyperbolic signature. By definition of spatial hypersurface the three tangent vectors $e^{\mu}_i$  ($e^{\mu}_i=\frac{\d u^{\mu}}{\d x^i}$) and the normal vector $\eta^{\mu}$ must satisfy the following orthogonal and completeness (\cite{thiemann}) relations:

\beq
\left\{
\begin{array}{l}
\eta^{\mu}\eta^{\nu}g_{\munu}=1\\
\eta^{\mu}e^{\nu}_ig_{\munu}=0\\
e^{\mu}_ie^{\nu}_jg_{\munu}= - \teta_{ij}
\end{array}
\right.
\label{relortonormal}
\eeq

\beq
\eta^{\mu}\eta^{\nu}-\teta^{ij}e^{\mu}_ie^{\nu}_j=g^{\munu}
\label{completeness}
\eeq
Clearly $\teta_{ij}$ is positive defined and represents the induced metrics on the spatial manifold. Now we can define $x^0$ as a time parameter and associate to every value of $x^0$ a different spatial manifold; then let $x^0$ vary with continuity: we get a  family of  hypersurfaces defined by the general parameterization 
\beq
u^{\mu}=u^{\mu}(x^i;x^0)
\label{par}
\eeq
 that represents a complete transformation of coordinates. The "time-deformation" vector $e^{\mu}_0=\frac{\d u^{\mu}}{\d x^o}$ links points with same spatial coordinates on two surfaces separated by an infinitesimal distance $dx_0$. By completeness this vector can be expressed in terms of the tangent and normal vector as follows
 \beq
 e^{\mu}_0=N\eta^{\mu} + S^ie^{\mu}_i
 \label{ezero}
 \eeq
 The components of the time deformation vector in this picture are called respectively \textit{Lapse} and \textit{Shift} functions and play a fundamental role in the ADM splitting. Finally, by using (\ref{ezero}) and (\ref{relortonormal}), we can recast the tensor metric according to the diffeomorphism (\ref{par}) and we get
\beq
g_{\munu}{}\Rightarrow{}\left(
\begin{array}{ccc}
N^2-\teta_{ij}S^iS^j & -\teta_{ij}S^j \\
                 &            \\
-\teta_{ij}S^j       & -\teta_{ij}
\end{array}
\right)
\label{admmetric}
\eeq
We can consider also the splitting for the inverse metrics:
\beq
g^{\munu}{}\Rightarrow{}\left(
\begin{array}{ccc}
\frac{1}{N^2} & -\frac{S^i}{N^2} \\
              &                  \\
-\frac{S^i}{N^2} & \frac{S^iS^j}{N^2}-\teta^{ij}
\end{array}
\right)
\quad\quad i,j=1,2,3
\label{adminversa}
\eeq
It can be shown that $N, S_i, \teta_{ij}$, are respectively scalar, vector and tensor under pure spatial diffeomorphisms and this is, indeed, the most important feature of this picture. This is known as the ADM splitting of the metrics, and it shows how we can redefine the ten degrees of freedom of the metrics $g_{\munu}$ in terms of objects with well-defined properties of transformation under spatial diffeomorphisms.

The advantage of the ADM splitting is clear when we consider the hamiltonian formulation of the dynamics.
In fact, a remarkable feature of the Einstein-Hilbert action is the presence of second time-derivatives of the metrics. This does not affect the lagrangian formulation that leads to Einstein's equation because , in the variational calculus, the term containing second time-derivatives  is only a global surface term, not really dynamic, but it is a problem in the hamiltonian approach, when we have to  calculate conjugates momenta of the metrics. So we need a method to identify , in the hamiltonian approach, the non dynamic part of the action. This is provided by the ADM splitting; the difference between the two approaches relies in the properties of Lagrangian and Hamiltonian under transformation of coordinates. While the former is manifestly covariant, the latter is covariant only for pure spatial transformations (due to definition of conjugates momenta in terms of first time derivatives); so we need to recast our degrees of freedom in order to identify correctly spatial or temporal objects, and this is obtained by ADM splitting.

Now, if we want to to consider the E-H action in terms of ADM variables we need the splitting of the curvature scalar. This is gained by the Gauss-Codacci formula (that comes from the splitting of Riemann tensor according to (\ref{admmetric})  (\cite{thiemann},\cite{wald},\cite{kolb}):
\beq
R=(K^2-K_{ij}K^{ij}-{}^3\!R)+2\nabla_{\nu}(\eta^{\mu}\nabla_{\mu}\eta^{\nu}-\eta^{\nu}K)
\label{gausscodacci}
\eeq
In this formula ${}^3\!R$ is the curvature scalar related to the spatial hypersurface (depending only on $\teta_{ij}$ and its spatial derivatives) and  $K_{ij}$ is the extrinsic curvature. We have:
\beq
K_{ij}=\frac{1}{2N}(D_iS_j+D_jS_i-\d_0\teta_{ij} )
\eeq
where $D_i$ is the covariant derivative defined on the spatial surface (then is $\teta_{ij}$-compatible).
The lowering/raising  that appears in $K^{ij}$, in its trace $K$, and in $N_i$ is of course made by $\teta_{ij}$. The other term that appears in Gauss-Codacci is not yet expressed in ADM variables but it doesn't matter because, when considering the action, it becomes a global surface term.
 Finally, by observing that we have $\sqrt{-g}=N\sqrt{\teta}$, we can give an expression for the Lagrangian in ADM variables:
\beq
L=bN\sqrt{\teta} (K^2-K_{ij}K^{ij}-{}^3\!R) +S.T.
\eeq
where $b=-\frac{1}{16\pi G}$.
As we can see , now the time derivative appears only at first order , so we can easily compute the conjugate momenta.
Moreover, we have no time derivative for \textit{Lapse} and \textit{Shift} functions; this means that they are not true dynamical variables and that the real dynamics is carried out only by the six degrees of freedom of the spatial metrics $\teta_{ij}$. So for the conjugate momenta ($\pi$ and $\pi^i$ for $N, S_i$ and $\pi^{ij}$ for $\teta_{ij}$ ) and their inverse relations we have:
\beq
\left\{
\begin{array}{l}
\pi^{ij}=-b\sqrt{\teta}(K\teta^{ij}-K^{ij} ) \\
\\
\dot{\teta}^{ij}=(D_iS_j+D_jS_i)-2NK_{ij} \\
\\
\pi\equiv\pi^i\equiv 0
\end{array}
\right.
\eeq

It can be shown that we can put the Hamiltonian in the following form:
\beq
H=NH^N+S_iH^i
\eeq
$H^N$ and $H^i$, respectively called Super-hamiltonian and Super-momenta, does not depend on the \textit{Lapse} or \textit{Shift} functions, as we can see by their explicitly expressions:

\beq
\left\{
\begin{array}{l}
H^N=-\frac{1}{b}G_{ijkl}\pi^{ij}\pi^{kl}+b\sqrt{\teta}{}^3\!R \\
\\
H^i=-2D_j\pi^{ij}
\end{array}
\right.
\eeq
$G_{ijkl}$ is called Supermetrics and have the following expression:
$$
G_{ijkl}=\frac{1}{2\sqrt{\teta}}(\teta_{ik}\teta_{jl}+\teta_{il}\teta_{jk}-\teta_{ij}\teta_{kl} )
$$
Hence, from a mathematical point of view, the  ADM splitting allows us to avoid the problem of second time derivatives. Furthermore it shows a remarkable structure of the dynamics. 
Conjugate momenta to \textit{Lapse} and \textit{Shift} are identically zero and so their time derivatives vanish; so , by examining the Poisson's  brackets we can see that the dynamics is constrained by the following relations:
$$
H^N\equiv H^i\equiv 0
$$
Then $N$ and $S_i$ play a role as lagrangian multipliers. This is a direct consequence of diffeomorphism invariance of theory and is , indeed, the most interesting feature  shown by ADM splitting (from the point of view of classical dynamics). Let's talk now about canonical quantization; it can be shown that, classically, the dynamics can be described simply by the first of these constrained equations; so, in order to have quantum dynamics we have to upgrade this equation to a quantum one, replacing conjugates momenta with appropriate derivatives operators. We have such an equation (Wheeler-DeWitt) (\cite{dewitt}) :
$$
\hat{H}^N\Psi=0
$$
Unfortunately, as an effect of the constraints, the equation we found is not evolutionary so its interpretation is not quite simple (\cite{isham},\cite{montaniwkbtempo},\cite{dewittempo},\cite{kolb}). The real problem is the meaning of time at quantum level and this is, indeed, still a matter of debate. A valuable approach to the WdW equation is its reformulation in terms of Ashtekar variables that lead to loop quantum gravity; another possible way is to consider the introduction of matter terms, from which we are able to obtain an unfrozen formalism (\cite{rovelli1},\cite{rovelli2},\cite{mercuri1},\cite{mercuri2}).

%AAAAAAAAAAAAAAAAAAAAAAAAAAAAAAAAAAAAAAAAAAAAAAAAAAAAAAAAAAAAAAAAAAAAAAAAAAAAAAAAAAAAAAAAAAAAAAAAAAAAAAAAAAAAAAAAAAAAAAAAAAAAAAAAAAAAAAAAAAAAAAAAAAAAAAAAAAAAAAAAAAAAAAAAAAAAAAAAAAAAAAA
\section{ADM Approach to Kaluza-Klein Model}
\label{tre}
Let's do some preliminaries observations.
While from a physical point of view KK reduction and ADM reformulation have a different meaning, however, from a mathematical point of view they both consist in a splitting of the metrics. Then, the ADM reformulation of KK model relies in performing these two splittings  starting from a 5-D manifold. At a first glance it seems that we only have to impose KK conditions and develop the ADM procedure taking into account the extra (space-like) dimension. Clearly, we have two ways, depending on which kind of splitting, KK or ADM, we want to perform as first step. It's useful to examine what we have to expect from these procedures.
Let's suppose to impose  KK condition first ; this leads to usual KK model, as we've seen. Now we can perform the usual $(3+1)$ ADM splitting of the 4-D metrics we have in this model and of the gauge-vector. However, in this way the space-time slicing is not complete because the extra dimension is not included in the splitting.
Then let's do as first the ADM splitting; this  corresponds to a (4+1) splitting and leads us to a 4-D spatial metrics, which take into account the extra dimension and to four \textit{Shift} functions; one of them is again related to the extra dimension. As a second step of this procedure we impose the KK condition and examine the KK reduction of the spatial metrics. By this procedure we find only a 3-D spatial gauge-vector and the dynamics lacks an explicit time component for the gauge vector , so we cannot be sure that this procedure, where the space-time slicing is complete, restores the correct KK model. 
Hence, we meet the real problems. First: are these two procedures equivalent ? Or, in other words, does KK reduction commute with ADM splitting ? Although this could appear only as a mathematical problem the question is not without physical relevance. In fact , as we've stressed in section (\ref{due}) KK reduction implies a breaking in the symmetry of the 5-D space ; so this could be not consistent with the hypotheses on which ADM splitting is constructed and the two procedures could lead to different dynamics. Second: it seems that both  procedures lead to some unsatisfactory features. Hence, what is really in trial now is the whole consistency of the application of ADM splitting to KK model or, more strictly,. the consistency of KK model itself. 
Now we'll examine in details the two procedures, starting with the splitting of the metrics.
\subsection{KK-ADM metrics}
Let's consider a 5-D tensor metric $J_{AB}$ (A,B=0,1,2,3,5). As seen in section (\ref{due}) (eq. \ref{splittingKK})  after KK reduction the metrics reads as follows:
$$
J_{AB}{}\Rightarrow{}\left(
\begin{array}{ccc}
g_{\munu}-\2f(ek)^2A_{\mu}A_{\nu} & -\2f(ek)A_{\mu} \\
                 &            \\
 & \\
-\2f(ek)A_{\mu}     & -\2f
\end{array}
\right)
$$
We have now to recast $g_{\munu}$ and $A_{\mu}$ in terms of ADM variables. For the space-time metrics this is easily done by using (\ref{admmetric}), so now we care about the gauge vector.
We have to notice that generally speaking the spatial components of any tensor don't correspond exactly to the correct spatial part of the tensor itself; the equivalence holds, as we'll see,  only in the covariant component picture. Indeed, for a spatial vector ${}^3\!A^i$ , defined on the spatial surface, we have ${}^3A^i=\teta^{ij}{}^3A_j$, whereas for the spatial components of a generic vector we have $A^i=g^{i\mu}A_{\mu}$ and so these components depend also on $A_0$ and on  \textit{Shift} functions; this means that although $\{A^i\}$ is a spatial vector (i.e.the components $A^i$ transform like a vector under pure spatial diffeomorphisms), it is not the complete  spatial part of the vector itself  (the exact equivalence hold only when the time deformation vector is normal to the surface, i.e. a synchronous reference) 
The most general way to split any vector or tensor according to ADM rules is to use the projection tensor, implicitly defined by the completeness relation (\ref{completeness})

\beq
\left\{
\bear{cc}
\eta^{\mu}\eta^{\nu}+q^{\munu}=g^{\munu} \\
\\
q^{\munu}=-\teta^{ij}e^{\mu}_ie^{\nu}_j
\ear
\right.
\label{rulpro}
\eeq
It's clear that $q^{\munu}$ acts as a projector on the spatial hypersurface and the spatial metrics itself is simply the projection tensor  recast in the spatial coordinates (\cite{thiemann}).
Thus, given a generic vector $B^{\mu}$ we can immediately obtain its spatial part $A^{\mu}_{\S}$ (i.e. $\eta_{\mu}A^{\mu}_{\S}\equiv 0$) from its contraction with the projection tensor:
$$
A^{\mu}_{\S} =q^{\mu}_{\nu}A^{\nu}
$$
 Then, in order to perform the ADM splitting we only have to recast the spatial part of the vector in the spatial 's coordinates picture; we have:
\beq
\left\{
\begin{array}{l}
{}^3A^i=\frac{\d{x^i}}{\d{u^{\mu}}}A^{\mu}_{\S} \\
{} \\
{}^3A_i=\frac{\d{u^{\mu}}}{\d{x^i}}g_{\munu}A^{\nu}_{\S}=\frac{\d{u^{\mu}}}{\d{x^i}} A_{\mu}=A_i
\end{array}
\right.
\label{ciccio}
\eeq
The equivalence that arises from the second of these relations is due to the definition  of tensor projection itself and to the  orthogonal relations (\ref{relortonormal}). Also, from the first of (\ref{ciccio}), we could show the following:
\beq
\left\{
\begin{array}{l}
A^0=\frac{1}{N^2}(A_0 -{}^3A_iS^i) \\
\\
A^i=-{}^3A^i-S^iA^0
\end{array}
\right.
\quad\quad\quad i,j=1,2,3
\label{acov}
\eeq
We can obtain the same result by simply considering a generic vector in covariant picture and its contraction with  the inverse metrics $g^{\munu}$ recast in term of ADM variables (see \ref{adminversa}).

Now let's turn to our task:
using (\ref{admmetric}) and observing that we are dealing with a gauge-vector in a covariant picture, we have the final splitting of the metrics :
\\
\\

$$J_{AB}\Rightarrow $$
\beq
\left(
\begin{array}{ccc}
 N^2-S_iS^i-(ek)^2\2f A_0^2   &\quad   -S_i-(ek)^2\2f A_0A_i      &\quad -ek\2f A_0 \\
\\
\\
-S_i-(ek)^2\2f A_0A_i         &\quad -\teta_{ij}-(ek)^2\2fA_iA_j  &\quad -ek\2f A_i \\
\\
\\
-ek\2f A_0                    &\quad -ek\2f A_i                   &\quad -\2f
\end{array}
\right)
\label{metricaKKADM}
\eeq 
where $A_i={}^3A_i$.
What about transformation rules? Transformations we are interested in are pure spatial KK diffeomorphism:
\beq
\left\{
\begin{array}{l}
x^4={x^4}'+ek\Psi({x^i}') \\
x^0={x^0}' \\
x^i= x^i ({x^i}')
\end{array}
\right.
\quad\quad\quad\quad i,j=1,2,3
\label{spazialiammissibili}
\eeq
$N$ ,$S_i$ and $\teta_{ij}$ come from $g_{\munu}$ that is a tensor without gauge component, so they are scalar, vector and tensor. $A_{\mu}$ presents a gauge-component, and so  $A_i$ does, while for pure spatial transformations $A_0$ is a scalar.
\beq
\left\{
\begin{array}{l}
{A_0}'=A_0 \\
\\
{A_i}'=A_j\frac{\d x^j}{\d {x^i}'} +\frac{\d \Psi}{\d {x^i}'} \quad\quad\quad (A_i={}^3\!A_i) \\
\\
{S_i}'=S_j\frac{\d x^j}{\d {x^i}'}
\end{array}
\right.
\quad\quad\quad i,j=1,2,3
\label{cov}
\eeq
Let's examine also the inverse metrics:
the KK reduction for the inverse metrics gives, as it can be easily verified by a direct calculus from (\ref{splittingKK}), the result
$$
J^{AB}\Rightarrow
\left(
\begin{array}{ccc}
g^{\munu} &\quad\quad -ekA^{\mu} \\
\\
\\
-ekA^{\mu} &\quad\quad (ek)^2A_{\mu}A^{\mu} -\frac{1}{\2f} 
\end{array}
\right)
\quad\quad \mu,\nu=0,1,2,3
$$
where $A^{\mu}=g^{\munu}A_{\nu}$.
Applying the ADM splitting, according to (\ref{adminversa}) we finally get:
\beq
J^{AB}\Rightarrow
\left(
\begin{array}{ccc}
\frac{1}{N^2} &\quad\quad  -\frac{S^i}{N^2}  &\quad\quad -ekA^0 \\
\\
\\
 -\frac{S^i}{N^2} &\quad\quad \frac{S^iS^j}{N^2} -\teta^{ij} &\quad\quad -ekA^i \\
\\
\\
-ekA^0 &\quad\quad -ekA^i &\quad\quad (ek)^2 (N^2{A^0}^2-A_iA_j\teta^{ij}) -\frac{1}{\2f}
\end{array}
\right)
\label{inversaKKADM}
\eeq
where now $A^i$ and $A^0$ are given by (\ref{acov}).
But now we have to notice that, according to (\ref{acov}), $A^0$  is not a  scalar and $A^i$ not a  gauge-vector. We have:
\beq
\left\{
\begin{array}{l}
{A^0}'=A^0-\frac{S^i}{N^2}\frac{\d \Psi}{\d x^i} \\
\\
{A^i}'=A^j \frac{\d {x^i}'}{\d x^j} +\frac{\d {x^i}'}{\d x^j}(\frac{S^iS^k}{N^2}-\teta^{jk})\frac{\d \Psi}{\d x^j} \quad\quad (A^i\neq {}^3\!A^i) \\
\\
{S^i}'=S^j\frac{\d {x^i}'}{\d x^j}
\end{array}
\right.
\quad\quad i,j=1,2,3
\label{contro}
\eeq

It' s now clear, from the step we've followed , that the space-time slicing is not complete because we don't take into account the extra-dimension; we only have performed  the splitting of the ordinary 4-D space-time and of the 4-D gauge vector; in other words we only have  defined a 3-d spatial hypersurface , not including the extra dimension . As we'll see, this will involve second time derivatives in the lagrangian formulation of the dynamics.
However, before studying the Lagrangian, we'll see the metrics that arises from the other procedure.

%AAAAAAAAAAAAAAAAAAAAAAAAAAAAAAAAAAAAAAAAAAAAAAAAAAAAAAAAAAAAAAAAAAAAAAAAAAAAAAAAAAAAAAAAAAAAA
\subsection{ADM-KK metrics}

Now we start with a $(4+1)$ ADM splitting of the 5-D metric tensor. So we split  our manifold in a direct product of a 4-d spatial manifold , including extra dimension, and of a one dimensional time-like manifold. Thus we get a 4-D spatial metrics $h_{\i\j}$ and four \textit{Shift} functions ($N_{\i}$).
\beq
J_{AB}{}\Rightarrow{}\left(
\begin{array}{ccc}
N^2-N^{\i}N_{\i}& -N_{\i} \\
                 &            \\
-N_{\i}       & -h_{\i\j}
\end{array}
\right)
\quad\quad\quad\quad \i,\j=1,2,3,5
\label{adm4+1}
\eeq
The next step is to examine the effects of KK reduction. Allowable (spatial) diffeomorphisms are simply:
\beq
\left\{
\begin{array}{l}
x^5={x^5}'+ek\Psi({x^i}') \\
 \\
x^i= x^i ({x^i}')
\end{array}
\right.
\quad\quad\quad\quad i,j=1,2,3
\label{spazialiammissibili2}
\eeq
Spatial metrics (positively defined) is split in a similar way to the complete 5-D metrics (\ref{splittingKK}) , with only a signature  change:
\beq
\left\{
\begin{array}{l}
h_{55}=\2f  \\
h_{5i}= \2f(ek) A_{i} \\
h_{ij}=\teta_{ij}+\2f(ek)^2A_{i}A_{j}
\end{array}
\right.
\quad\quad\quad\quad i,j=1,2,3
\label{losplittingKK4+1}
\eeq
In the above formulas $\phi$ is still a scalar field , $A_i$ is a 3-D spatial gauge vector and $\teta_{ij}$ is the 3-D spatial metrics. 
It's also useful to examine the 4-bein picture of this metrics, which indeed, is similar (excerpt for the dimension) to the usual KK basis set:

\beq
\left\{
\begin{array}{l}
e^{(5)}_{\i}=\phi (ekA_i, 1) \ \ \ \ \ \ \ \ \ \\
\\
e^{(j)}_{\i}= (u^{(j)}_{i}, 0 )
\end{array}
\right.
\label{comptetrade4+1}
\eeq

\beq
\left\{
\begin{array}{l}
e^{\i}_{(5)}= (0,0,0,\frac{1}{\phi}) \\
\\
e^{\i}_{(j)}= (u^i_{(j)}, -ekA_{i}u^{i}_{(j)})
\end{array}
\right.
\label{comptetradeinv4+1}
\eeq
$$
 \quad (\i),(\j)=(1),(2),(3),(5) \quad\quad (i),(j)=(1),(2),(3)
$$
$$
\i,\j= 1,2,3,5
$$
$$
i,j=1,2,3
$$
Clearly $u^i_{(j)}$ and its inverse define the 3-D spatial metrics $\teta_{ij}$ and have no gauge components in its transformation rules.
Finally, we have:
\bitem
\item splitting ADM-KK

\beq
\left(
\begin{array}{ccc}
N^2-h_{\i\j}N^{\i}N^{\j} &\quad\quad -N_{i} &\quad\quad -N_5 \\
\\
\\
-N_{i} &\quad\quad -\teta_{ij}-(ek)^2\2f  A_iA_j &\quad\quad -ek\2f A_i \\
\\
\\
-N_5 &\quad\quad -ek\2f A_{i} &\quad\quad -\2f
\end{array}
\right)
\label{metricaADMKK}
\eeq
where explicitly results
$$
h_{\i\j}N^{\i}N^{\j} = (\phi N^5)^2+2ek\2f A_{i}N^{i}N^5+\teta_{ij}N^{i}N^{j}+(ek)^2\2f A_iA_jN^iN^j
$$
In a same way we can also obtain the AD-KK splitting for the inverse metrics and we have:
\item splitting ADM-KK inverse metrics

\beq
\left(
\begin{array}{ccc}
\frac{1}{N^2} &\quad\quad -\frac{N^{i}}{N^2} &\quad\quad -\frac{N^5}{N^2} \\
\\
\\
-\frac{N^{i}}{N^2} &\quad\quad -\teta^{ij}-\frac{N^{i}N^{j}}{N^2} &\quad\quad \frac{N^{i}N^5}{N^2}+ekA_{j}\teta^{ij} \\
\\
\\
-\frac{N^5}{N^2} &\quad\quad \frac{N^{i}N^5}{N^2}+ekA_j\teta^{ij} &\quad\quad \frac{{N^5}^2}{N^2}-\frac{1}{\2f}-(ek)^2A_{i}A_j\teta^{ij}
\end{array}
\right)
\label{inversaADMKK}
\eeq
\eitem

The most evident feature of this procedure is , as we've announced, that this picture lacks an explicit time component for the gauge vector.
However, it has to be said, our analysis of the effects of KK conditions is not yet finished. In fact the controvariant components $N^{\i}$ that appear in the inverse metrics are still linked to covariant ones by the 4-D metrics $h_{\i\j}$ as it arises from the $(4+1)$ splitting of the inverse metrics. So we need to identify the correct 3-D spatial part of \textit{Shift} functions whose indexes must be lowered/raised with the 3-D spatial metrics $\teta_{ij}$ that is the only metrics we have to take into account after KK splitting. 

 In order to solve this problem let us consider the \textit{Shift} functions before the symmetry breaking. $N_{\i}$ and $N^{\i}$ transform as 4-D spatial vector. By imposing the KK restriction we have the following transformation rules:
\beq
\left\{
\begin{array}{l}
{N_{i}}'=ekN_5\frac{\d \Psi}{\d {x^{i}}'}+N_{j}\frac{\d x^{j}}{\d {x^{i}}'} \\
\\
{N_5}'=N_5 
\end{array}
\right.
\label{trasfNbasso}
\eeq

\beq
\left\{
\begin{array}{l}
{N^{i}}'=N^{j}\frac{\d {x^{i}}'}{\d x^{j}} \quad\quad\\
\\
{N^5}'=N^5-ekN^{j}\frac{\d\Psi}{\d x^{j}} 
\end{array}
\right.
\label{trasfNalto} 
\eeq
Then we see that $N_5$ is a scalar and $N^i$ a vector, while in the other components appear gauge-terms that must be related to a dependence on the gauge-vector.
In order to examine in more details the structure of the \textit{Shift} functions, we can express them in the 4-bein picture. It results:
\beq
\left\{
\begin{array}{l}
N_5=N_{(5)}\phi \\
\\
N_{i}=N_{(l)}u^{(l)}_{i}+ekN_5A_{i}
\end{array}
\right.
\eeq
\beq
\left\{
\begin{array}{l}
N^5=\frac{N^{(5)}}{\phi}-ekA_{i}N^{(l)}u_{(l)}^{i} \\
\\
N^{i}=N^{(l)}u_{(l)}^{i}
\end{array}
\right.
\eeq
This picture  clearly  agrees with  the transformation rules, but,  it also reveals  how \textit{Shift} functions contains a  gravitational (pure vectorial) part and a gauge one.
We define as $S_i$ the vectorial part of $N_i$, and by its definition we have $S^i=\teta^{ij}S_j$, so we can rewrite:
\beq
\left\{
\begin{array}{l}
N_5=N_{(5)}\phi \\
\\
N_{i}=S_{i}+ekN_4A_{i}
\end{array}
\right.
\label{exptetradebassa2}
\eeq
\beq
\left\{
\begin{array}{l}
N^5=\frac{N^{(5)}}{\phi}-ekA_{i}S^{i} \\
\\
N^{i}=S^{i}
\end{array}
\right.
\label{exptetradealta2}
\eeq

Now we make a comparison between the ADM-KK and KK-ADM metrics, for the elements $J_{05}$ and $J_{0i}$. If the metrics are equivalent, the following expression must be satisfied.

\beq
\left\{
\begin{array}{l}
N_{i}=S_{i}+ (ek)^2\2fA_0A_{i} \\
\\
N_5=ek\2fA_0
\end{array}
\right.
\quad\quad i= 1,2,3
\label{conversione1}
\eeq
This suggests to define $A_0$ in terms of $N_5$ as in the second equation. Is this definition acceptable ?
Indeed, the second members of these equations verify the transformations rules (\ref{trasfNbasso}), as we can see by (\ref{cov}) and also the 4-bein structure. Moreover they lead also to an equivalence for the $J_{00}$ component of the metrics, so the (\ref{conversione1}), that we label "conversion formulas"  appear correct.
Then, if we accept as true the (\ref{conversione1}) we can say that vectorial part of $N_i$ corresponds to the 3-D  \textit{Shift} vector and we are able to identify a time-component for the gauge-vector. However, we have to be sure , in defining a 4-d gauge vector, that the (\ref{conversione1}) conversion formulas rebuild the particular relationships which link covariant and controvariant components, as we have seen in ADM splitting (see eq. \ref{acov}). 
So, we need to examine  the inverse metrics too. By comparison of the $J^{05}$ and $J^{0i}$ elements of the ADM-KK and KK-ADM metrics, we find the requirement:
\beq
ekA^0=\frac{N^5}{N^2}
\label{x}
\eeq
\beq
-ekA^{i}=\frac{N^{i}N^5}{N^2}+ek{}^3\!A^{i}
\quad\quad i= 1,2,3
\label{y}
\eeq
By using these equations with (\ref{conversione1}),(\ref{exptetradebassa2}), (\ref{exptetradealta2}) we find:
\beq
A^0=\frac{1}{N^2}(A_0-S^{i}A_{i})
\eeq
\beq
A^{i}=-S^{i}A^0-{}^3\!A^{i}
\eeq
Hence, conversion formulas rebuild exactly the ADM formulas for the splitting of a vector. Moreover, it can be shown that conversion formulas lead to the equivalence of all others elements of the metrics.

So, also in ADM-KK procedure ( that is the one with complete space-time separation ) we can rebuild a 4-D gauge vector. 
From a mathematical point of view we can say that the two metrics gained with the ADM-KK and KK-ADM procedure are equivalent; at the same way, from this point of view,  conversion formulas shows how, by KK condition, the 4-D \textit{Shift} vector is split into a pure vectorial part and in a gauge one. It is remarkable that $A_0$ is linked to the extra dimension \textit{Shift} component. We'll return on this point.
Anyway, to make us sure that these formulas have a physical meaning, we have to examine the ADM-KK and KK-ADM Lagrangian, and see if they commute. 

 Following expression summarize the conversion formulas:
\beq
\left\{
\begin{array}{l}
N_{i}=S_{i}+ (ek)^2\2fA_0A_{i} \\
\\
N_5=ek\2fA_0
\end{array}
\right.
\quad\quad\quad
\left\{
\begin{array}{l}
N^{i}=S^{i} \\
\\
N^5=ekN^2A^0
\end{array}
\right.
\label{conversione}
\eeq
\beq
S^{i}=\teta^{ij}S_{j}
\eeq

%AAAAAAAAAAAAAAAAAAAAAAAAAAAAAAAAAAAAAAAAAAAAAAAAAAAAAAAAAAAAAAAAAAAAAAAAAAAAAAAAAAAAAAAAAAAAAAAAAAAAAAAAAAAAAAAAAAAAAAAAAAAAAAAAAAAAAAAAAAAAAAAAAAAAAAAAAAAAAAAAAAAAAAAAAAAAAAAAAAAAAAAAAAAAAAAAAAAAAAAAAAAAAAAAAAAAAAAAAAAAAAAAAAAAAAAAAAAAAAAAAAAAAAAAAAAAAAAAAAAAAAAAAAAAAAAAAAAAA

\subsection{Lagrangian}

Conversion formulas allow us to compute ADM-KK and KK-ADM Lagrangian in the same set of dynamical variables, all with well-defined transformation properties under pure spatial KK diffeomorphisms.
\bitem
\item $\teta_{ij}$, tensor; $S_i$ vector; $N$, scalar
\item $A_0$, scalar; $A_i$ gauge-vector
\item $\phi$ scalar
\eitem

\subsubsection{KK-ADM Lagrangian}
\

Consider the action of the usual KK model:
$$
S=b \int\! d^4\!x \sqrt{-g}\quad(\phi R+2\nabla_{\mu}\d^{\mu}\phi+\frac{1}{4}(ek)^2\phi^3F_{\munu}F^{\munu})
\quad\quad b=-\frac{1}{16\pi G}
$$
Second term in brackets is a total surface term , so we can omit it. Now we have to make the ADM splitting of the electromagnetic term and the curvature term.
In order to do this we simply need the completeness relation (\ref{completeness}) and the projection rules for tensors that reads (see also \ref{rulpro})
$$
T^{\munu}_{\S}=q^{\mu}_{\rho}q^{\nu}_{\sigma}T^{\rho\sigma}
$$
\beq
\left\{
\begin{array}{l}
T^{ij}=\frac{\d{x^i}}{\d{u^{\mu}}}\frac{\d{x^j}}{\d{u^{\nu}}}T^{\munu}_{\S} \\
{} \\
T_{ij}=\frac{\d{u^{\mu}}}{\d{x^i}}\frac{\d{u^{\nu}}}{\d{x^j}}T_{\munu}
\end{array}
\right.
\label{admtensor}
\eeq
Hence, the electromagnetic term is split as follows:

$$
F_{\munu}F^{\munu}=F_{ij}{}^3\!F^{ij}-\frac{2}{N^2}\teta^{ij}M_iM_j
$$
where $M_i=F_{i0}-S^iF_{ij}$ and ${}^3\!F^{ij}$ is gained from $F_{ij}$ via $\teta^{ij}$.
For the curvature term we can employ the Gauss-Codacci formula (\ref{gausscodacci}), but now, because of the presence of $\phi$, we must take into account the term which formerly was a surface one.

$$
2b\sqrt{-g}\,\phi\nabla_{\nu}(\eta^{\mu}\nabla_{\mu}\eta^{\nu}-\eta^{\nu}K)
$$
In order to split this term we need again to use the projection tensor, but, by the presence of the covariant derivative, the projection rules is changed; it can be shown that the correct formula is the following: given a generic vector $B_{\nu}$ we have:
$$
q^{\munu}(\nabla_{\mu}B_{\nu}) = q^{\munu}D_{\mu}B_{\nu}+B^{\eta}K
$$
where $D_{\mu}$ is the 4-D covariant derivative compatible with $q_{\munu}$ that, when recast in the 3-D spatial picture, provides the 3-D covariant derivative constructed via $\teta_{ij} $, and $B^{\eta}$ reads  $B^{\eta}=B_{\mu}\eta^{\mu}$.
Thus, setting $B_{\mu}=\nabla_{\mu}\phi$ in the above formula , after some algebra we get our desired splitting of the extra gravitational term as follows

\beq
\phi\nabla_{\nu}(\eta^{\mu}\nabla_{\mu}\eta^{\nu}-\eta^{\nu}K)=-\d_{\eta}\d_{\eta}\phi +\teta^{ij}D_i\dj\phi + T.S
\label{splittingextratermine}
\eeq
where $D_i$ is the 3-D covariant derivatives and
$$
\d_{\eta}=\frac{1}{N}(\d_0-S^i\di)
$$

Finally , Lagrangian takes the following expression:
\begin{eqnarray}
\L_{KK-ADM}&=&b\sqrt{\teta}\,N\phi(K^2-K_{ij}K^{ij}-{}^3\!R3)+2b\sqrt{\teta}\,N(D_i\d^i\phi-\d_{\eta}\d_{\eta}\phi)+ \nonumber \\
&{}&+\frac{b}{4}\sqrt{\teta}\,N(ek)^2{\phi}^3(F_{ij}{}^{3}\!F^{ij}-\frac{2}{N^2}M_iM^i)
\label{lagKKADM}
\end{eqnarray}

\bitem
\item $b=\frac{-1}{16\pi G}$
\item $\d_{\eta}\phi=\frac{1}{N}(\d_0\phi-S^i\di\phi)$
\item $M_i=F_{i0}-S^jF_{ij}$
\item $K_{ij}=\frac{1}{2N}(D_iS_j+D_jS_i-\d_0\teta_{ij})$
\eitem
As we can see the term $\d_{\eta}\d_{\eta}\phi$ contains a second time derivative and this is due to a not complete space-time splitting. To stress this we can put:
\beq
\L_{KK-ADM}=\widetilde{\L} -2b\sqrt{\teta}\,N\d_{\eta}\d_{\eta}\phi
\label{KKADM1}
\eeq
with $\widetilde{\L}$ implicitly defined by (\ref{lagKKADM}).

\subsubsection{ADM-KK Lagrangian}

Now we use first ADM splitting to gain a complete $(4+1)$ space-time slicing. We start from the 5-d action and use the Gauss-Codacci formula, omitting the surface term. After this we impose KK conditions; by the cylindricity condition we make the dimensional reduction and finally action takes the following form:
$$
S=b \int\!d^4\!x N\sqrt{h} (\K^2-\K_{\i\j}K^{\i\j}-{}^4\!R) \quad\quad \i,\j=1,2,3,5
$$
where $h_{\i\j}$ is the 4-d spatial metrics we have seen formerly. ${}^4\!R$ and $\K$ are respectively the curvature scalar and the extrinsic curvature related to $h_{\i\j}$, and we need their KK splitting. This can be achieved via the 4-bein formalism and we will only give the results. For the intrinsic curvature term we have :
\beq
{}^4\!R={}^3\!R-\frac{2}{\phi}\teta^{ij}D_{i}\d_j\phi-\frac{1}{4}(ek)^2\2f F_{ij}{}^{3}\!F^{ij}
\quad\quad i,j=1,2,3
\eeq
where $\teta_{ij}$ is the 3-D spatial metrics, ${}^3\!R$ and $D_i$ are curvature scalar and covariant derivative related to $\teta_{ij}$ and $F_{ij}$ is the spatial tensor constructed with the 3-D gauge-vector $A_i$.
In a same way the splitting of extrinsic curvature terms give us the following result:
\beq
\K^2-\K_{\i\j}\K^{\i\j}=[K^2-K_{ij}K^{ij}]-\frac{2}{\2f}K\phi\d_{eta}\phi-\frac{1}{2N^2}\2fM_{i}M^{i}
\eeq
where $K_{ij}$ is the 3-D extrinsic curvature and $\d_{\eta}$, $M_i$ are the same objects we have defined in the KK-ADM procedure.
Finally, we have the following expression for the Lagrangian:
\begin{eqnarray}
\L_{ADM-KK}&=&b\sqrt{\teta}\,N\phi(K^2-K_{ij}K^{ij}-{}^3\!R)+2b\sqrt{\teta}\,N(D_{i}\d^{i}\phi-K\d_{\eta}\phi)+ \nonumber \\
&{}&+\frac{b}{4}\sqrt{\teta}\,N(ek)^2{\phi}^3(F_{ij}{}^{3}\!F^{ij}-\frac{2}{N^2}M_{i}M^{i})
\label{lagADM-KK}
\end{eqnarray}
It is worth noting that we have only a term containing a time derivative of the scalar field. We can rewrite the Lagrangian as follows:
\beq
\L_{ADM-KK}=\widetilde{\L}-2b\sqrt{\teta}\,NK\d_{\eta}\phi
\eeq
By comparison with the KK-ADM Lagrangian we can see how the term $\widetilde{\L}$ is the same in both procedures. Thus, let us examine the remaining terms with time derivatives of $\phi$.

Let us rewrite the terms that give a difference in the two Lagrangian:
\bitem
\item ADM-KK $\quad\quad -2b\sqrt{\teta}\,NK\d_{\eta}\phi$
\item KK-ADM $\quad\quad -2b\sqrt{\teta}\,N\d_{\eta}\d_{\eta}\phi$
\eitem
Now we will show how these terms are equivalent apart from a total surface term.
For a generic matrix $M(x)$ and its inverse we have the identity
\beq
Tr[M^{-1}(x)\di M(x)]=\di [ln(detM(x))]
\label{idut}
\eeq
By (\ref{idut}) we can express $K$ in a useful form;
\begin{eqnarray}
K=\teta_ij^{ij}K_{ij} &=& \frac{\teta^{ij}}{2N}(D_iS_j+D_jS_i-\d\teta_{ij}) \nonumber \\
               &=& \frac{1}{N}D_iS^i-\frac{1}{2N}\teta^{ij}\d_0\teta_{ij} \nonumber \\
               &=& \frac{1}{N}D_iS^i-\frac{1}{N}\frac{1}{\sqrt{\teta}}\d_0\sqrt{\teta}
\label{ppp}
\end{eqnarray}
and, by (\ref{ppp}), we can rewrite the ADM-KK term as follows:
\bitem
\item ADM-KK$\quad\quad N\sqrt{\teta}\,K\d_{\eta}\phi= \sqrt{\teta}\,(D_iS^i)(\d_{\eta}\phi)-(\d_o\sqrt{\teta})(\d_{\eta}\phi)$
\eitem
Now let us turn our attention to KK-ADM term
\beq
N\sqrt{\teta}\,\d_{\eta}\d_{\eta}\phi =\sqrt{\teta}\d_0\d_{\eta}\phi-\sqrt{\teta}S^i\di\d_{\eta}\phi 
\eeq
For the first term we have:
\beq
\sqrt{\teta}\d_0\d_{\eta}\phi=\d_o(\sqrt{\teta}\d_{\eta}\phi)-(\d_0\sqrt{\teta})(\d_{\eta}\phi)
\eeq
so we can observe how a surface term appears, by the presence of a total time derivative.
For the second term, observing that we have $\d_{\eta}\phi=\eta^{\mu}\dm\phi$ and so $D_i\d_{\eta}\phi=\di\d_{\eta}\phi$ we have:
\beq
\sqrt{\teta}S^i\di\d_{\eta}\phi=\sqrt{\teta}D_i(S^i\d_{\eta}\phi)-\sqrt{\teta}(D_iS^i)(\d_{\eta}\phi
\eeq
First term of the second member is a surface term, too, for Gauss theorem in curved space. Hence we have:

\bitem
\item KK-ADM $\quad\quad N\sqrt{\teta}\,\d_{\eta}\d_{\eta}\phi=\sqrt{\teta}\,(D_iS^i)(\d_{\eta}\phi)-(\d_o\sqrt{\teta})(\d_{\eta}\phi) +T.S $
\eitem
It follows that, apart from surface term, the two Lagrangians are equivalent, so that ADM-KK and KK-ADM procedures lead to the same dynamics. Hence , both procedures will lead to the same Hamiltonian. Moreover this provides a physical meaning for the conversion formulas.

\subsubsection{Hamiltoniam formulation}

Starting from Lagrangian we have provided , by Legendre's transformation, we can obtain the Hamiltonian for the KK 5-D model. It results that the conjugates momenta of $N,S_i$ and $A_0$ are identically zero and the real dynamics is carried out by the spatial metrics $\teta_{ij}$, as in the 4-d case, and by $A_i$ and $\phi$.
The Hamiltonian function reads :
\beq
\mathcal{H}=NH^{N}+S_iH^{i}+A_{0}H^0
\label{hamiltoniana}
\eeq
where we have:

\begin{eqnarray}
H^{N} &=& b\sqrt{\teta}\phi R-2b\sqrt{\teta}D^i\di\phi-\frac{1}{2b\sqrt{\teta}\phi}T_{ijkl}\Sigma^{ij}\Sigma^{kl}-\frac{1}{6b\sqrt{\teta}}\pi^2_{\phi}\phi+ \nonumber \\
    &{}& +\frac{1}{3b\sqrt{\teta}}\pi_{\phi}\Sigma^{ij}\teta_{ij}-\frac{1}{4}b\sqrt{\teta}(ek)^2\phi^3F_{ij}F^{ij}-\frac{2}{b\sqrt{\teta}(ek)^2\phi^3}\pi^i\pi^j\teta_{ij}
\label{hn}
\end{eqnarray}

\beq
H^i= -2D_j\Sigma^{ij}+\pi_{\phi}\d^{i}\phi-\pi^jF^{i\ }_{\ j}
\label{hi}
\eeq

\beq
H^0=-D_i\pi^i
\label{hao}
\eeq
where $\Sigma^{ij},\pi^i,\pi_{\phi}$ are respectively the conjugates momenta to $\teta_{ij},A_i,\phi$ and $T_{ijkl}$ is defined as follows:
$$
T_{ijkl}=(-\frac{2}{3}\teta_{ij}\teta_{kl}+\teta_{ik}\teta_{jl}+\teta_{il}\teta_{jk})
$$
As we have seen in the 4-D case, $N,S_i,A_0$ are not true dynamical variables but they play the role of Lagrange's multipliers, providing the following constraints for the dynamics:
$$
H^n\equiv H^i\equiv H^0\equiv 0
$$
These constraints can be used instead of the complete equations for the study of the dynamics.
Really, a remarkable feature of this Hamiltonian is given by the relation we have seen between $A_0$ and $N_5$.
In fact, before the symmetry breaking due to KK conditions, we have a Lagrangian and an Hamiltonian analogues the those of the 4-d theory, except for the dimension and the number of \textit{Shift} functions that provides constraint for the dynamics, due to the diffeomorphism invariance. After the symmetry breaking we have the electromagnetism coupling and the presence of the constraint related to $A_0$, due to gauge-invariance. So, the relation between $A_0$ and $N_5$ shows how the electromagnetic constraint arises as a particular case of the diffeomorphism invariance. Moreover this mean that the hamiltonian formalism we have developed holds at the same way before and after the symmetry breaking and can itself take into account the alteration provided by the KK conditions.

\section{The role of the cylindricity hypothesis}
\label{quattro}
Once  proved the consistency of the ADM reformulation of KK model, we now want to see in more details the role of the cylindricity hypothesis.  In this task we'll use the so-called "space-ambient embedding " picture of a curved manifold; so now we review briefly the main features of this technique.
Consider a Minkowskian vectorial space, $\mathcal{V}^{n+1}$, $n+1$-dimensional and its canonical basis of orthonormal vectors provided by the set $\{\n_i\}$;
\bitem
\item $\{\n_i\}$ \quad\quad\quad  i=0,1,...,n
\item $\n_i\cdot\n_j = \eta_{ij}$
\eitem
where $\eta_{ij}$ is the Minkowsky matrix.
In the following we will refer to $\mathcal{V}^{n+1}$ as the "Space-Ambient".
A generic manifold $\mathcal{M}^n$, n-dimensional and $C^{\infty}$, embedded in the Space-Ambient can be described by a parameterization such as
$$
\vett{Y}=\vett{Y}(u^{\alpha})=Y^i(u^{\alpha}\n_i
$$
where $\vett{Y}$ belongs to $\mathcal{V}^{n+1}$ and the $n$ parameters $u^{\alpha}  (\alpha= 0,1,...,n-1) $ define local coordinates on the manifold. For an assigned parameterization the hyperplane tangent on the manifold is defined , point  by point, through the $n$ tangent vectors
$$
\f_{\alpha}=\frac{\d\vett{Y}}{\d u^{\alpha}}=\frac{\d Y^i}{\d u^{\alpha}}\n_i
$$
We simply say that a vector belonging to $\mathcal{M}^n$ is a vector belonging to such a hyperplane. In this way we can employ the usual vectorial representation and have a local basis for the manifold.
Hence, for a manifold vector we have two possible representations:
\bitem
\item $\vett{V}=V^{\alpha}\f_{\alpha}$
\item $\vett{V}=V^i\n_i$
\eitem
where $V^i=V^{\alpha}\frac{\d Y^i}{\d u^{\alpha}}$
It can be shown that the components $V^{\alpha}$ are controvariant for a change of the parameterization (that is the same of a coordinates transformation). Moreover, we can define the components of the metric tensor of the manifold as the scalar products of local basis vectors
\beq
g_{\alphabeta}=\f_{\alpha}\cdot\f_{\beta}
\label{scalarmetric}
\eeq
The covariant component of a vector can also be defined in terms of scalar product and we have
\beq
V_{\alpha}=g_{\alphabeta}V^{\beta}=\vett{V}\cdot\f_{\alpha}
\label{scalarcovariant}
\eeq
Within this picture we can easily rebuild all the typical relationships of general relativity like covariant derivation rules and so on. The most relevant advantage we take of this picture is that it allows us to keep a vectorial representation also in curved space.
Hence, let's consider now a 5-d manifold, embedded in a 6-D Minkowskian space-ambient, with a set of five basis vectors $\{\f_{\mu},\f_5\}$. 

In the ADM-KK procedure we perform a complete space-time slicing providing four space-like vectors and a normal time-like one, plus a time deformation vectors:
\beq
\left\{
\bear{l}
\f_0=N\ev+N_i\f_i + N_5\f_5 \\
\ev\cdot\f_i\equiv 0 \\
\ev\cdot\f_5\equiv 0
\ear
\right.
\eeq

In the KK-ADM procedure we first perform KK reduction; hence, according to
 (\ref{scalarmetric}), the basis vectors must satisfy the following relations:
 \beq
\left\{
\begin{array}{l}
\f_5\cdot\f_5=-\2f \\
\\
\f_5\cdot\f_{\mu}=-(ek)\2f A_{\mu} \\
\\
\f_{\mu}\cdot\f_{\nu}= g_{\munu}-(ek)^2\2f A_{\mu}A_{\nu}
\end{array}
\right.
\label{baseKK2}
\eeq
Moreover, according to  KK allowable diffeomorphisms, we have the following transformation rule:
\beq
\left\{
\begin{array}{l}
{\f_5}'=\f_5 \\
\\
{\f_{\mu}}'=\f_{\mu}\frac{\d x^{\mu}}{\d {x^{\mu}}'} +\frac{d\psi}{\d {x^{\mu}}'}
\end{array}
\right .
\label{trasfbaseKK}
\eeq
At this point, in KK-ADM procedure we have to perform the splitting of the 4-D space-time starting  by the four vector $\f_{\mu}$ and defining a new set of three space-like vectors and its normal time-like one;

$$
\{\f_{\mu}\}\rightarrow \{\e_i,\ev\}
$$
\beq
\left\{
\bear{l}
\e_0=N\ev+S_i\e_i  \\
\ev\cdot\e_i\equiv 0 
\ear
\right.
\eeq
In this case we have in general $\f_5\ev\neq 0$. So, at a first glance, in our procedures we have two different relations of orthogonality between $\ev$ and $\f_5$; these conditions are scalar, and, furthermore,  as we see in the first of (\ref{trasfbaseKK}), $f_5$ sets a preferred direction in space, as a direct consequence of the cylindricity hypothesis. Finally we have also two different definitions of the time deformation vectors that could lead to different time variables and time derivatives.
Thus, the space-ambient picture shows in a striking way the problems that made us  doubt about the consistency of ADM reformulation.
But if we continue our analysis we can see that in the KK-ADM procedure, as we presented it at this point, the ADM splitting is not well defined:
 we turn our attention to $\e_i$;
$$
\e_i=e^{\mu}_i\f_{\mu}
$$
As we can see by transformation rules (\ref{trasfbaseKK}) a change in the local coordinates also change the basis vector , because of the presence of the gauge-terms.
 In order to have consistency we cannot do the splitting by starting from the vectors $\f_{\mu}$.
Consider now the third equation of the (\ref{baseKK2}); we define two new sets of vectors satisfying the following relations:. 
\bitem
\item $\f_{\mu}=\vett{u}_{\mu}+\vett{a}_{\mu}$
\item $\vett{u}_{\mu}\cdot\vett{a}_{\nu}\equiv 0$
\item $\vett{u}_{\mu}\cdot\vett{u}_{\nu}=g_{\munu} $
\item $\vett{a}_{\mu}\cdot\vett{a}_{\nu}=-\2fA_{\mu}A_{\nu}$
\eitem
These definitions  still agree with the third equation of (\ref{baseKK2}) but now
the vectors $\vett{u}_{\mu}$ doesn't have gauge-components. 
By this re-definition for $\f_{\mu}$, consider now the second of (\ref{baseKK2}).
We can assume as true the condition
\beq
\f_5\cdot\vett{u}_{\mu}\equiv 0
\label{quondam}
\eeq
Indeed, if this condition were not true we would have a pure vectorial term that, really, is not present in our equation.
The set $\{\vett{u}_{\mu}\}$ defines moreover the real space-time  metrics so that it is correct to do with this set the ADM splitting. If we define $\ev$ from this set we have,  by  (\ref{quondam})    the condition $\ev\cdot\f_5\equiv 0$.
Hence, while the cylindricity conditions sets the existence of a preferred direction,   is the cylindricity itself that always sets  $\f_5\cdot\ev=0$. Moreover, as a final remark, let's examine the  time deformation vector and the time derivative that arise from it in both procedure; we have:
\bitem
\item ADM-KK $\f_0=N\ev+N_i\f_i + N_5\f_5 \Rightarrow \d_0=N\d{\eta}+N^i\di + N^5\d_5$
\item KK-ADM $\e_0=N\ev+S_i\e_i \Rightarrow \d_0=N\d{\eta}+S^i\d_i $
\eitem
 As we can see, reminding that from conversion formulas we have $N^i=S^i$  and observing that from cylindricity $\d_5$ has no effect , we have the equivalence of the two definitions.

\section{Final remarks}
\label{last}

Now let us summarize the results of this paper. The main part of our work was devoted to check the consistency of the ADM reformulation of the KK model or, at the same way, as we've seen, to check the commutation of KK reduction with the ADM procedure. We've stressed the hypotheses on which they are constructed and that the physics underlying this two splitting is of different kind  ; and we've discuss how this could make us  doubt about the commutation of the two procedures. So we've studied in details the metrics and the relative dynamics resulting in both case (ADM-KK and KK-ADM); we have strictly shown how it's possible to provide conversion formulas that make the metrics to be the same and lead to an equivalent dynamics.  Lagrangians are the same apart from surface term and both can be recast in the same set of variables; more important, we can always choose the set in which appears the time component of the gauge field. Indeed, as result from the conversion formulas the time component of the gauge field is proportional to the fifth \textit{Shift} function; moreover this condition is sufficient, by the general properties of \textit{Shift} functions, to conclude  that $A_0$ will correctly appear as a lagrange multipliers in the Hamiltonian.
At this point we are able to take two conclusions.

 First, the proof of the commutation and the above results about the time component of the gauge field, are a positive check for the self-consistency of the KK model ; in this perspective we've also seen the role of the cylindricity condition . Indeed this result shows in more detail how the gauge invariance arises from the diffeomorphism invariance; in fact, although we have a symmetry breaking provided by KK reduction, we still keep having as many constraints as the number of spatial dimensions; we have a change in the physical meaning of the constraint that arises from the extra dimension: from a geometric invariance to a gauge one. This enforces the idea that KK model is not simply an artifice  but has a real physical meaning.

Second, these results are also a check for the utility of the hamiltonian formulation. In fact, let's figure a point in the time, during the evolution of the 5-D universe , in which  h KK symmetry breaking happens. In order to describe physics at quantum scale (or at least to attempt to do it) we need the hamiltonian formulation; before of this point the hamiltonina formulation  is carried out with the $4+1$ splitting and no way KK splitting is admitted; the last one can hold only after symmetry breaking, so this leads us to the ADM-KK procedure which , as we've seen lacks the time component of the gauge field. But, via the conversion formulas that start to hold after symmetry breaking, we can recast dynamic variables in a suitable set containing the $A_0$ component. Hence we can say that the Hamiltonian holds at every time, is well-defined in a unique way and is able to take into account the changes due to the symmetry breaking.
This kind of result has to be viewed as a first step in the study of the quantum properties of the model. This analysis, which we purpose to do prosecuting this work, can be viewed both in the spirit of the usual canonical quantization and in the path of the Ashtekar reformulation and loop quantum gravity.

\section*{Acknowledgements}
We are grateful to Orchidea Maria Lecian for improving the English of this paper.
%\section*{References}

\end{document}